\documentstyle[preprint,aps,prb,epsfig]{revtex}
\begin{document}
\draft
\pagestyle{plain}
\newcommand{\D}{\displaystyle}
\title{\bf Weak anisotropic impurity scattering 
in unconventional superconductors}
\author{Grzegorz Hara\'n\cite{AA} and A. D. S. Nagi}
\vspace{0.4cm}
\address{Department of Physics,
University of Waterloo,
Waterloo, Ontario,
Canada, N2L 3G1}
\date{July 24, 1996} 
\maketitle

\begin{abstract}
The effect of weak anisotropic (momentum-dependent) impurity scattering 
in unconventional superconductors has been investigated. 
It is shown that the anisotropic scattering can lead either to 
a small reduction or a small enhancement of the isotropic pair-breaking effect. 
The influence of the anisotropy of the scattering potential becomes significant  
for the order parameters with large Fermi surface average values. 
In that case an unexpected enhancement (up to $\sim 10\%$) of the  
critical temperature $T_c$ over the critical temperature in the absence 
of impurities $T_{c_{0}}$ is predicted for a small impurity concentration.\\ 
\end{abstract}
\vspace{0.5cm}
\noindent
\pacs{Keywords: A. superconductors C. point defects D. phase transitions} 

\vspace{1cm}
\noindent
{\it Reprinted from Solid State Communications, Vol. 101, No. 1,\\ 
G. Hara\'n and A. D. S. Nagi, 
Weak anisotropic impurity scattering 
in unconventional superconductors, pp. 71-75, 
1997, with permission from Elsevier Science Ltd, The Boulevard, 
Langford Lane, Kidlington 0X5 1GB, UK}

\newpage
There has been a great interest in studying the properties of unconventional 
superconductivity in last decade. These superconductors are characterized  
by order parameter of lower symmetry than the conventional one,  
which only breaks U(1) gauge symmetry. Such order parameters are involved 
in almost all theories of  
heavy fermion and high temperature superconductivity. \cite{14,25,26,27}    
As the order parameter $\Delta\left({\bf k}\right)$ is a nontrivial real function 
of ${\bf k}$ its Fermi surface average $\left<\Delta\left({\bf k}\right)\right>$  
strongly depends on the shape of the Fermi sheet and may take values between 
-1 and 1 under the normalization condition 
$\left<\Delta^2\left({\bf k}\right)\right>=1$. Most of the theoretical effort has 
been focussed on the case of $\left<\Delta\left({\bf k}\right)\right>=0$ however, 
and only recently the unconventional superconductivity with a nonvanishing 
Fermi surface average of the order parameter has been discussed. 
\cite{33,34,35,30,24,3,4} In the presence of impurities particularly, these 
superconductors were suggested to show some interesting phenomena in the density       
of states, which should detectable in the thermodynamic and transport properties. 
\cite{33,34,35,24,3,4} 

In this paper, we go beyond the approximation of the isotropic 
impurity scattering and consider the weak anisotropic (momentum-dependent) 
impurity potential, which can be treated as a small perturbartion of   
the isotropic scattering. We show   
that a really interesting case of weak anisotropic scattering is
when the Fermi surface average of the order parameter is nonzero. Particularly
for large values of $\left<\Delta\left({\bf k}\right)\right>$ we obtain
a surprizing  increase of the critical temperature $T_c$ with the impurity
scattering rate in the range of small impurity concentration.

Recent experimental \cite{11,21} as well 
as theoretical \cite{1} studies suggest the importance of the anisotropic 
impurity scattering issue in the 
interpretation of the $T_c$ suppression due to impurities in YBCO. As   
the anisotropy of the impurity potential is preasumably strong   
in this compund \cite{21,1} the use of a perturbation method may not   
be applicable. We believe, however, that the results of the present  
study may apply to some other anisotropic superconductors like heavy fermions,   
for instance, where a highly complex Fermi surface \cite{28} (FS) and   
anisotropic order parameter may lead to a nonzero  
$\left<\Delta\left({\bf k}\right)\right>$. \cite{31,29} 
We take $\hbar=k_{B}=1$ throughout the paper.\\

We assume randomly distributed nonmagnetic impurities in an anisotropic
superconductor. Treating the electron-impurity scattering within
first-order Born approximation and neglecting the impurity-impurity
interaction, \cite{2} the normal and anomalous temperature Green's
functions averaged over the impurity positions are given by

\begin{equation}
\label{e2}
G\left(\omega,{\bf k}\right)=
-\frac{i\tilde{\omega}+\xi_{k}}
{\tilde{\omega}^{2}+{\xi_{k}}^{2}+|\tilde{\Delta}\left(
{\bf k}\right)|^{2}}
\end{equation}

\begin{equation}
\label{e3}
F\left(\omega,{\bf k}\right)=
\frac{\tilde{\Delta}\left({\bf k}\right)}
{\tilde{\omega}^{2}+{\xi_{k}}^{2}+|\tilde{\Delta}\left(
{\bf k}\right)|^{2}}
\end{equation}

\noindent
where the renormalized Matsubara frequency
$\tilde{\omega}\left({\bf k}\right)$ and the renormalized
order parameter $\tilde{\Delta}\left({\bf k}\right)$ read 

\begin{equation}
\label{e4}
\tilde{\omega}\left({\bf k}\right)=\omega+i
n_{i}\int |w\left({\bf k}-{\bf k'}\right)|^{2}
G\left(\omega,{\bf k'}\right)\frac{d^{3}k'}{\left(2\pi\right)^{3}}
\end{equation}

\begin{equation}
\label{e4a}
\tilde{\Delta}\left({\bf k}\right)=\Delta\left({\bf k}\right)+
n_{i}\int |w\left({\bf k}-{\bf k'}\right)|^{2}
F\left(\omega,{\bf k'}\right)\frac{d^{3}k'}{\left(2\pi\right)^{3}}
\end{equation}

\noindent
In above $\omega=\pi T(2n+1)$ (T is the temperature and n is an integer number),
$\xi_{k}$ is the quasiparticle energy, $n_{i}$ is impurity (defect)
concentration, $w\left({\bf k}-{\bf k'}\right)$ is momentum dependent impurity
potential and $\Delta\left({\bf k}\right)$ is the orbital part of a singlet
\cite{12} superconducting order parameter defined as

\begin{equation}
\label{e1}
\Delta\!\left({\bf k}\right)=\Delta e\!\left({\bf k}\right)
\end{equation}

\noindent
where $e\left({\bf k}\right)$ is a real basis function of an one dimensional 
(1D) irreducible representation of a crystal point group or a linear combination 
of the basis functions of 1D representations. 
We normalize $e\!\left({\bf k}\right)$ by taking 
$\left<e^{2}\right>=1$, 
where $<...>=\int_{FS}dS_k n\left({\bf k}\right)\left(...\right)$ 
denotes the average over the Fermi surface (FS), 
$n\left({\bf k}\right)$ is the angle resolved FS density 
of states normalized to unity, i.e. $\int_{FS}dS_k n\left({\bf k}\right)=1$,  
and $\int_{FS}dS_{k}$ stands for the integration over the Fermi surface.\\

The impurity scattering potential is assumed to be separable and given by

\begin{equation}
\label{e19}
|w\left({\bf k}-{\bf k'}\right)|^{2}=|w|^{2}
h\left({\bf k}\right)h\left({\bf k'}\right)
\end{equation}

\noindent
with

\begin{equation}
\label{e19a}
h\left({\bf k}\right)=1+g\left({\bf k}\right),\;\;\;\;\;
|g\left({\bf k}\right)|\ll 1
\end{equation}

\noindent
which means that $g\left({\bf k}\right)$ function represents a small anisotropic
correction to the isotropic scattering potential. The above functions are 
normalized by taking $\left<h\right>=1$ ($\left<g\right>=0$). 
We note from Eq. (\ref{e4}) and the form of impurity potential 
(Eqs. (\ref{e19}) and (\ref{e19a})) that $\tilde{\omega}$ is ${\bf k}$-dependent.
This means that the electron self-energy due to impurity scattering 
and consequently the quasiparticle life-time are anisotropic and change 
over the Fermi surface. Further, it yields from Eqs. (\ref{e4a}),    
(\ref{e19}), and (\ref{e19a})  that the impurity scattering may change the 
symmetry of the renormalized order parameter $\tilde{\Delta}\left({\bf k}\right)$  
depending on the $g\left({\bf k}\right)$ symmetry.  In that respect our
approximation differs from that by  Markowitz and Kadanoff \hspace{1em}\cite{6} 
as well as that by Millis et al. \cite{17} where the authors   
assumed only a change of a degree of order parameter anisotropy but not
the anisotropy function itself.\\ 

To proceed further, we restrict the wave vectors of the electron self-energy
and pairing potential to the Fermi surface and
replace $\int d^{3}k/\left(2\pi\right)^{3}$
by $N_{0}\int_{FS}dS_{k}n\left({\bf k}\right)\int d\xi_{k}$,
where $N_{0}$ is the overall density of states at the Fermi surface. 
Using Eqs. (\ref{e2}), (\ref{e3}), (\ref{e1}), and (\ref{e19})  
in Eqs. (\ref{e4}) and (\ref{e4a}) and performing the integration
over $\xi_{k}$ (particle-hole symmetry of quasiparticle spectrum is assumed)
we write

\begin{eqnarray}
\label{e9}
\tilde{\omega}\left({\bf k}\right) & = &
\omega\left[1+u\left(\omega\right)h\left({\bf k}\right)\right]\\
\tilde{\Delta}\left({\bf k}\right) & = & \Delta\left[e\left({\bf k}\right)
+e\left(\omega\right)h\left({\bf k}\right)\right]
\end{eqnarray}

\noindent
where $u\left(\omega\right)$ and $e\left(\omega\right)$ functions are 
determined by the self-consistent equations 

\begin{eqnarray}
\label{e10a}
u\left(\omega\right) & = & \Gamma\int_{FS}dS_{k}n\left({\bf k}\right)
h\left({\bf k}\right)\frac{1+u\left(\omega\right)h\left({\bf k}\right)}
{\left[\tilde{\omega}^{2}+
|\tilde{\Delta}\left({\bf k}\right)|^{2}\right]^{1/2}}
\end{eqnarray}

\begin{eqnarray}
\label{e10c}
e\left(\omega\right) &= & \Gamma\int_{FS}dS_{k}n\left({\bf k}\right)
h\left({\bf k}\right)\frac{e\left({\bf k}\right)+e\left(\omega\right)
h\left({\bf k}\right)}
{\left[\tilde{\omega}^{2}+
|\tilde{\Delta}\left({\bf k}\right)|^{2}\right]^{1/2}}
\end{eqnarray}

\noindent
and $\Gamma=\pi N_{0}n_{i}|w|^{2}$ is the impurity scattering rate.
The gap function is given by the self-consistent equation 

\begin{equation}
\label{e15}
\Delta\left({\bf k}\right)=-T\sum_{\omega}\sum_{{\bf k'}}
V\left({\bf k}, {\bf k'}\right)
\frac{\tilde{\Delta}\left({\bf k'}\right)}
{\tilde{\omega}^{2}+{\xi_{k'}}^{2}+|\tilde{\Delta}\left(
{\bf k'}\right)|^{2}}
\end{equation}

\noindent
where $V\left({\bf k},{\bf k'}\right)$ is the phenomenological pair potential
taken as

\begin{equation}
\label{e15a}
V\left({\bf k},{\bf k'}\right)=-V_{0}e\left({\bf k}\right)
e\left({\bf k'}\right)
\end{equation}

\noindent
Following standard procedure, \cite{13} we obtain the equation for
the critical temperature $T_{c}$ as

\begin{equation}
\label{e15b}
\ln\frac{T_{c}}{T_{c_{0}}}=2\pi T_{c}\sum_{\omega>0}
\left[\left(f\left(\omega\right)\right)_{\Delta=0}-\frac{1}{\omega}\right]
\end{equation}

\noindent
with

\begin{equation}
\label{e15c}
\left(f\left(\omega\right)\right)_{\Delta=0}=
\D\int_{FS}dS_{k}n\left({\bf k}\right)
\frac{e\left({\bf k}\right)}{\tilde{\omega}_{0}\left({\bf k}\right)}
\left[\frac{\tilde{\Delta}\left({\bf k}\right)}
{\Delta}\right]_{\Delta=0}
\end{equation}

\noindent
where $T_{c_{0}}$ is the critical temperature in the absence of impurities and
\mbox{$\tilde{\omega}_{0}\left({\bf k}\right)=
\tilde{\omega}\left({\bf k}\right)_{\Delta=0}$}.
Using Eqs. (\ref{e19a})-(\ref{e10c}) and neglecting terms of order 
$g^{2}\left({\bf k}\right)$ and higher, we obtain for $\Delta\rightarrow 0$ 

\begin{equation}
\label{e23}
\tilde{\omega}_{0}\left({\bf k}\right)=\omega+\Gamma
\left(1+g\left({\bf k}\right)\right)sign\left(\omega\right)
\end{equation}

\begin{equation}
\label{e24}
\left[\frac{\tilde{\Delta}\left({\bf k}\right)}{\Delta}\right]_{\Delta=0}=
e\left({\bf k}\right)+
\left(\left<e\right>+
\left<eg\right>\left(1-\frac{\Gamma}
{|\omega|+\Gamma}\right)\right)
\frac{\Gamma}{|\omega|}\left[1+g\left({\bf k}\right)\right]
\end{equation}

\noindent
We note from Eq. (\ref{e24}) that the impurity scattering induces anisotropy in  
the renormalized order parameter $\tilde{\Delta}\left({\bf k}\right)$ other than 
that of $\Delta\left({\bf k}\right)$ which is described by a function 
$e\left({\bf k}\right)$ (Eq. (\ref{e1})). There are two different sources of 
this additional anisotropy in $\tilde{\Delta}\left({\bf k}\right)$. First of 
them is a nonzero value of $\left<e\right>$ which is determined by the symmetry 
of the order parameter and that of the Fermi surface, and the second reflected 
in Eq. (\ref{e24}) by $g\left({\bf k}\right)$ function and coupling coefficient 
$\left<eg\right>$ is the anisotropy of the impurity scattering potential.\\  

Based on Eqs. (\ref{e15c}), (\ref{e23}), and (\ref{e24}) we obtain from 
Eq. (\ref{e15b}) 

\begin{equation}
\label{e25}
\begin{array}{l}
\D\ln\frac{\D T_{c}}{\D T_{c_{0}}}=\left(\left<e\right>^{2}-1\right)
\left(\psi\left(\frac{\D 1}{\D 2}+\frac{\D\Gamma}{\D 2\pi T_{c}}\right)
-\psi\left(\frac{\D 1}{\D 2}\right)\right)+\\
\\
\D\left(2\left<e\right>\left<eg\right>-\left<e^{2}g\right>\right)
\frac{\D\Gamma}{\D 2\pi T_{c}}\psi^{(1)}
\left(\frac{\D 1}{\D 2}+\frac{\D\Gamma}{\D 2\pi T_{c}}\right)
\end{array}
\end{equation}

\noindent
where $\psi\left(z\right)$ and $\psi^{(1)}\left(z\right)$ are digamma  
and trigamma functions respectively.  
The first term on the righthand side of Eq. (\ref{e25}) represents the 
isotropic impurity scattering effect on $T_{c}$ \cite{9,15,32} and the  
influence of the weak anisotropic scattering is reflected by the second term.  
Compared to the isotropic pair-breaking, the weak anisotropic impurity 
potential can lead either to a small (of order $g$) additional decrease 
of $T_{c}$ when the term $(2\left<e\right>\left<eg\right>-\left<e^{2}g\right>)$ 
is negative or to a small reduction of $T_{c}$ suppression for a positive 
value of $(2\left<e\right>\left<eg\right>-\left<e^{2}g\right>)$. 

We consider the case of a positive value of this parameter in more detail. 
For the sake of convenience we assume $\left<e^{2}g\right>=0$ and then 
deal with the order parameter-impurity potential coupling coefficient 
$\left<eg\right>$, which does not reduce the generality of approach but 
simplifies the notation. One would obtain the same results without the 
above assumption using the term $(2\left<e\right>\left<eg\right>- 
\left<e^{2}g\right>)$ as a parameter with the values assigned to 
$2\left<e\right>\left<eg\right>$ coefficient in this paper. The actual value of 
$\left<e^{2}g\right>$ depends on two factors: the symmetry of both 
$e\left({\bf k}\right)$, $g\left({\bf k}\right)$ functions and the shape 
of the Fermi surface. Thus the estimation of $\left<e^{2}g\right>$ parameter 
is difficult in the real systems, however, its value may vanish in some  
cases (see Appendix). 
   
We present the critical temperature $T_{c}$ normalized by the 
critical temperature in the absence of impurities $T_{c_{0}}$ as a function of 
normalized impurity scattering rate $\Gamma/2\pi T_{c_{0}}$ for different 
values of the impurity coupling coefficient $\left<eg\right>$ in Figs. 1-3. 
As we are considering a weak anisotropic potential and have neglected terms 
of order $\left<eg\right>^2$ while writing Eq. (\ref{e25}), we take values 
of $\left<eg\right>$ upto $\sim 0.3$. 
We notice from these figures that the effect of anisotropic scattering is 
very weak for small 
Fermi surface average values of the order parameter $\left<e\right>$ and may not 
be distinguished from the isotropic impurity scattering ( Figs. 1-2), nevertheless 
the influence of the anisotropy in the scattering potential increases with the  
increase in $\left<e\right>$. For the large values of the Fermi surface 
averaged order parameter 
that is $\left<e\right>\sim 0.9$  (Fig. 3) and for the coupling coefficient 
$\left<eg\right>$ larger than approximately 0.1 we observe a very interesting 
feature of an initial enhancement of $T_c$ for small impurity concentration.  
This initial enhancement of $T_c$ over $T_{c_{0}}$ may be understood by 
writing Eq. (\ref{e25}) for small impurity concentration (i.e. 
$\Gamma/2\pi T_{c_{0}}\ll 1$). We have in that case  

\begin{equation}
\label{e26} 
\D\frac{T_c}{T_{c_{0}}}-1\simeq\frac{\pi^2\alpha}{2}
\left(\frac{\Gamma}{2\pi T_{c_{0}}}\right)
\end{equation}

\noindent 
where 
$\alpha=\left<e\right>^2-1+2\left<e\right>\left<eg\right>-\left<e^2g\right>$. 
Therefore $T_c$ is an increasing function of $\Gamma$ for a positive $\alpha$ 
and decreases with the impurity concentration for a negative $\alpha$. 
This criterion  
may serve for estimation of $\left<eg\right>$ value which gives rise to the 
enhancement of the critical temperature with scattering rate. For example  
for $\left<e\right>=0.8$ ($\left<e^2g\right>=0$) this coefficient is 
$\left<eg\right>\sim 0.23$, and for $\left<e\right>=0.9$ 
($\left<e^2g\right>=0$) the coefficient $\left<eg\right>=0.2$ leads to 
the $T_c$ enhancement of about $10\%$ for $\Gamma/2\pi T_{c_{0}}\simeq 0.12$ 
(as in Fig. 3). The required large $\left<e\right>$ value suggests a 
substantial s-wave component in the order parameter.  

Although in above we have taken $\left<e^2g\right>=0$, 
we expect, that even when $\left<e^2g\right>$ coefficient has a positive value,  
the factor $\alpha$ can be made positive 
by appropriately large value of $\left<e\right>$. It is worth mentioning that 
the above results cannot be obtained with the assumption 
of the proportionality of the impurity potential anisotropy function  
$g\left({\bf k}\right)$ to the order parameter orbital 
function $e\left({\bf k}\right)$, since the constraint $\left<g\right>=0$ 
would yield $\left<e\right>=0$.\\ 

In summary, we have investigated the effect of weak anisotropic impurity 
scattering in unconventional superconductors. It is shown that 
the anisotropy of the impurity potential can lead to either a small  
increase or a small decrease in $T_c$ suppression due to isotropic 
scattering. The influence of weak anisotropic scattering becomes significant 
for the order parameters with large Fermi surface average values 
$\left<e\right>$. In this case the critical temperature is increased  
up to 10 per cent over $T_{c_{0}}$ for small impurity concentration. 
This anisotropic scattering-induced $T_c$ enhancement in  
unconventional superconductors with large 
$\left<e\right>$ values is a novel feature of the present study.\\ 

This work was supported by the Natural Sciences and Engineering 
Research Council of Canada. 

\appendix  
\section*{}   
We assume $e\left({\bf k}\right)$  and $g\left({\bf k}\right)$ as the basis  
functions of 1D irreducible 
representations of the crystal point group. Therefore for a symmetry operation 
from this group $S_i$   

\begin{equation}
\label{a1} 
S_i g\left({\bf k}\right)=g_i g\left({\bf k}\right)
\end{equation}

\noindent
where $g_i$ is a number. Further, if there is a part $P$ of FS   
which reproduces the whole Fermi sheet under the crystal group symmetry  
operations (FS=$\sum_i S_i P$) and 
the average value of $g\left({\bf k}\right)$ over this segment of FS  
does not vanish, i.e. $\left<g\right>_P\neq 0$, then from the normalization 
$\left<g\right>=0$ and from the relation 

\begin{equation}
\label{a2}
\left<g\right>=\sum_i \left<S_i g\right>_P=\left<g\right>_P\sum_i g_i
\end{equation}

\noindent
we get 

\begin{equation}
\label{a3}
\sum_i g_i=0
\end{equation}

\noindent
Taking into account that $e^2\left({\bf k}\right)$ 
is an identity representation for a real $e\left({\bf k}\right)$, 
which means that $S_i e^2\left({\bf k}\right)=e^2\left({\bf k}\right)$,  
we calculate $\left<e^2g\right>$ as follows 

\begin{equation}
\label{a4}
\left<e^2g\right>=\sum_i\left<S_i\left(e^2g\right)\right>_P=
\left<e^2g\right>_P\sum_i g_i 
\end{equation}

\noindent
Then Eqs. (A3) and (A4) yield $\left<e^2g\right>=0$.  

On the other hand, if the order parameter is given by a linear combination 
of a s-wave component and a basis function $f\left({\bf k}\right)$ of a 
nonidentity 1D irreducible representation of the crystal point group, that is 

\begin{equation}
\label{a5}
e\left({\bf k}\right)=\left<e\right>+f\left({\bf k}\right)
\end{equation}

\noindent
then $\left<eg\right>=\left<fg\right>$ and under the same assumptions 
about $g\left({\bf k}\right)$ we get 
$\left<e^2g\right>=2\left<e\right>\left<eg\right>$ which leads to a  
cancelation of the anisotropy of the scattering potential in 
$T_c$ equation (Eq. (\ref{e25})). 
We conclude from this, that in order to observe the effect of weak  
anisotropic scattering, $e\left({\bf k}\right)$ and $g\left({\bf k}\right)$ 
must belong to $A_{1g}$ representation with 
$g\left({\bf k}\right)$ normalized to fulfill $\left<g\right>=0$. 

The above considerations, however, are based on the assumption that  
$g\left({\bf k}\right)$ transforms according to the symmetry of the 
crystal point group and are not valid  
in the case of $g\left({\bf k}\right)$ described by a symmetry 
other than that of the crystal lattice.

\newpage
\begin{center}
\begin{figure}[p]
\parbox{0.1cm}{\large\vfill $$T_c/T_{c_{0}}$$\vspace{3ex}\vfill }
\parbox{15cm}{\epsfig{file=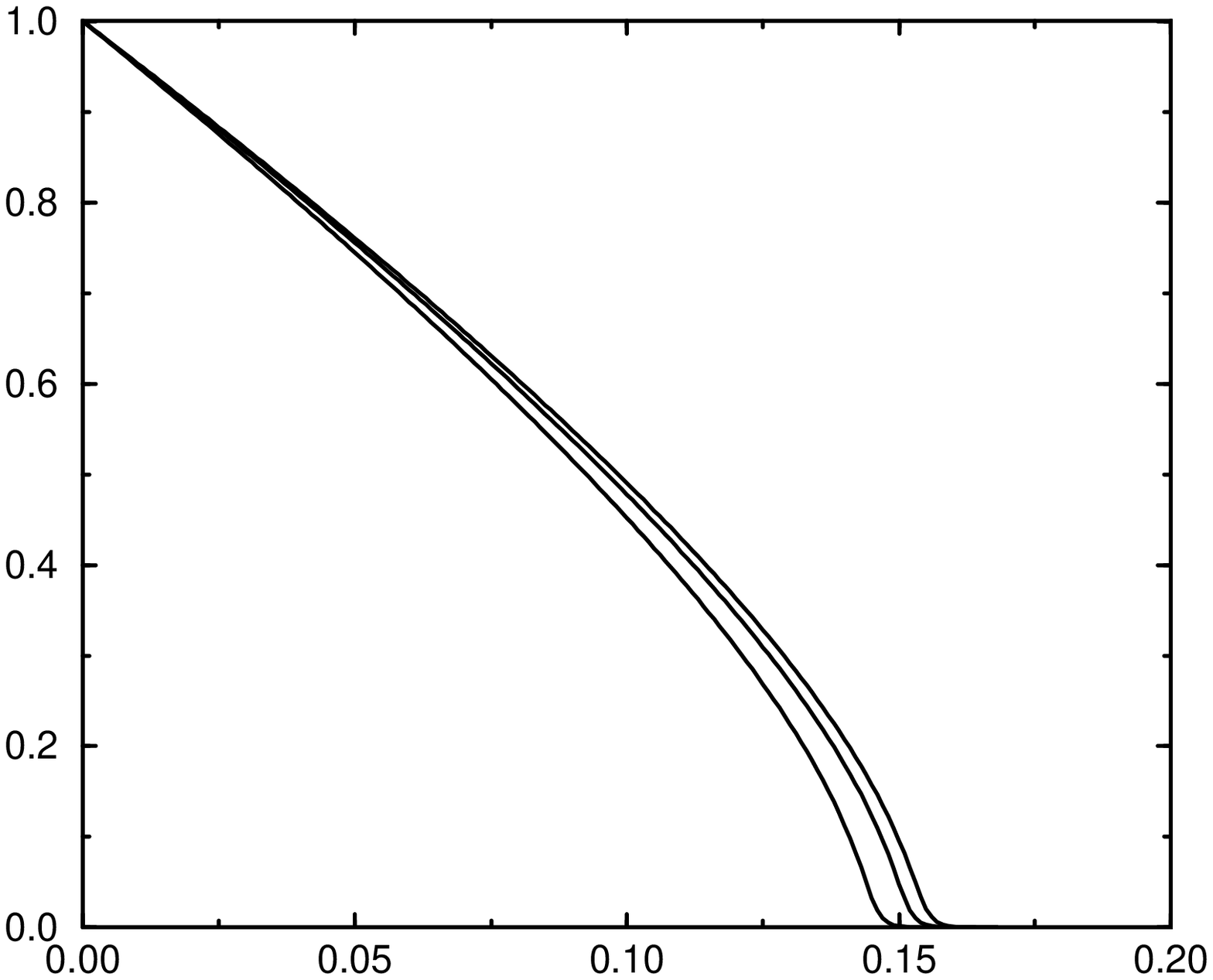,height=15cm,width=15cm} }
\parbox{0.5cm}{\hfill} 
\parbox{18cm}{\large\vspace{-9ex}\hfill $$\Gamma/(2\pi T_{c_{0}})\;\;\;\;\;\;$$\hfill}
\caption{Normalized critical temperature $T_c/T_{c_{0}}$
as a function of the normalized scattering rate 
$\Gamma/2\pi T_{c_{0}}$ for $\left<e\right>=0.1$. 
From the bottom to the top, curves are plotted for 
$\left<eg\right>=$0 (isotropic scattering), 0.2, and 0.3; 
$\left<e^{2}g\right>=0$ is assumed.}  
\end{figure}
\end{center}

\newpage
\begin{center}
\begin{figure}[p]
\parbox{0.1cm}{\large\vfill $$T_c/T_{c_{0}}$$\vspace{3ex}\vfill }
\parbox{15cm}{\epsfig{file=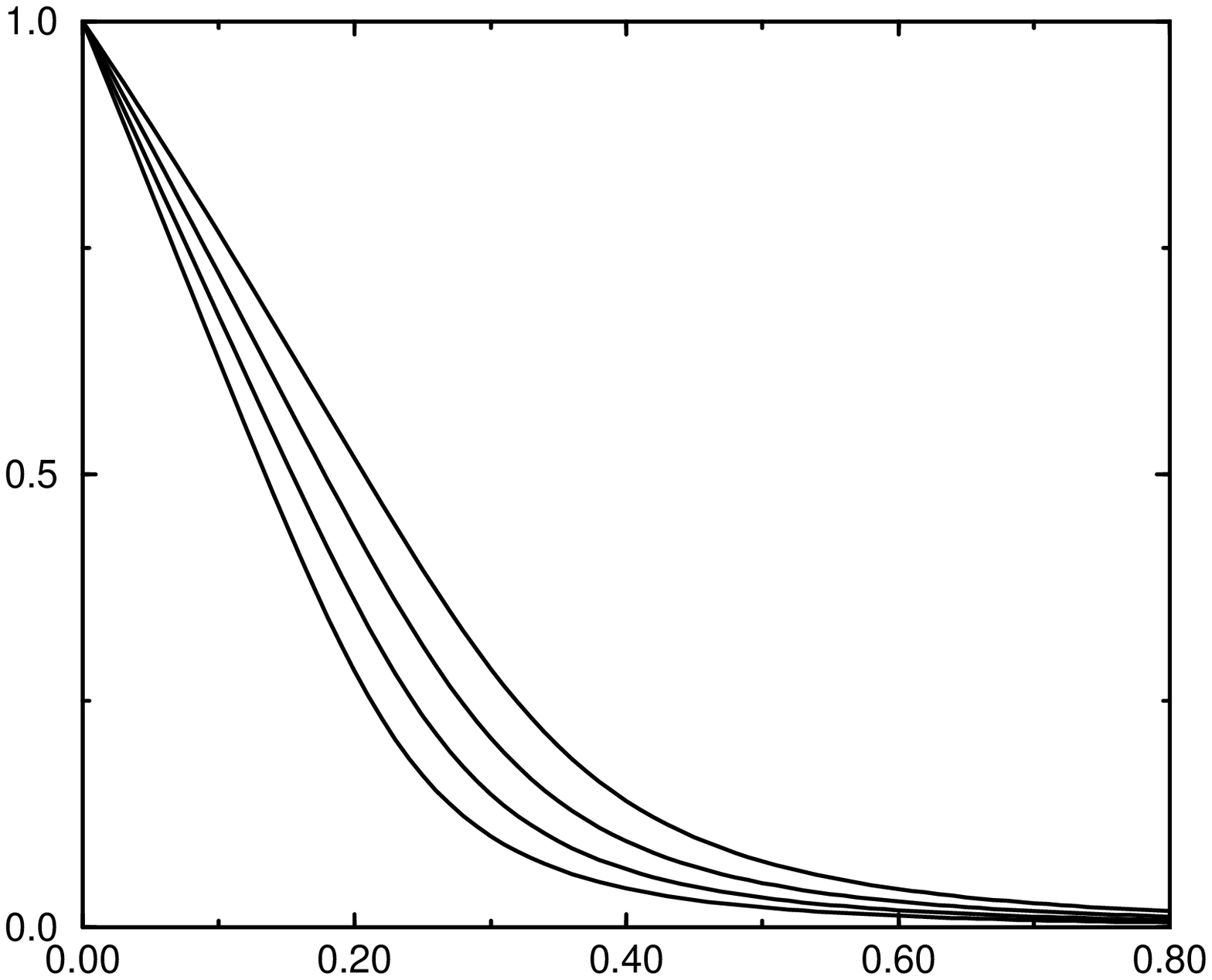,height=15cm,width=15cm} }
\parbox{0.5cm}{\hfill}
\parbox{18cm}{\large\vspace{-9ex}\hfill $$\Gamma/(2\pi T_{c_{0}})\;\;\;\;\;\;$$\hfill}
\caption{Normalized critical temperature $T_c/T_{c_{0}}$
as a function of the normalized scattering rate 
$\Gamma/2\pi T_{c_{0}}$ for $\left<e\right>=0.5$. 
From the bottom to the top, curves are plotted for 
$\left<eg\right>=$0 (isotropic scattering), 0.1, 0.2, and 0.3; 
$\left<e^{2}g\right>=0$ is assumed.}
\end{figure}
\end{center}

\newpage
\begin{center}
\begin{figure}[p]
\parbox{0.1cm}{\large\vfill $$T_c/T_{c_{0}}$$\vspace{3ex}\vfill }
\parbox{15cm}{\epsfig{file=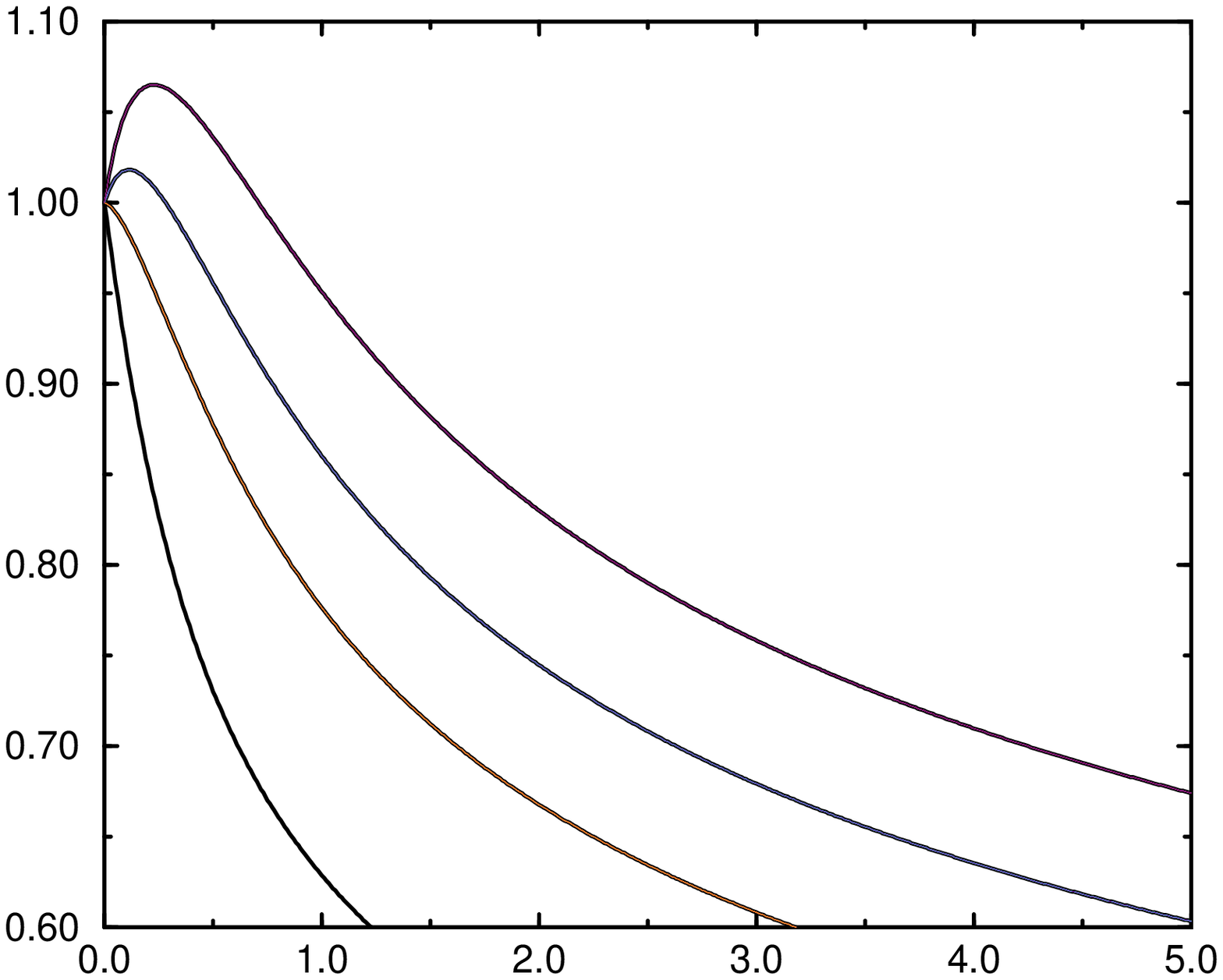,height=15cm,width=15cm} }
\parbox{0.5cm}{\hfill}
\parbox{18cm}{\large\vspace{-9ex}\hfill $$\Gamma/(2\pi T_{c_{0}})\;\;\;\;\;\;$$\hfill}
\caption{Normalized critical temperature $T_c/T_{c_{0}}$
as a function of the normalized scattering rate 
$\Gamma/2\pi T_{c_{0}}$ for $\left<e\right>=0.9$. 
From the bottom to the top, curves are plotted for 
$\left<eg\right>=$0 (isotropic scattering), 0.1, 0.15, and 0.2; 
$\left<e^{2}g\right>=0$ is assumed.}
\end{figure}
\end{center}


\begin{references}
\bibitem{12} Our discussion is also valid for a triplet-paired
state since the nonmagnetic impurities do not affect
the spin part of the order parameter.
\end{references}
\end{document}